\documentclass[aps,prl,reprint,superscriptaddress,10pt]{revtex4-2}

\usepackage{amsmath,amssymb,mathrsfs}
\usepackage{psfrag}
\usepackage{graphicx}
\usepackage{graphics}
\usepackage{epsfig}
\usepackage{bm}
\usepackage{color}
\usepackage{verbatim,color}
\usepackage{physics}
\usepackage[normalem]{ulem}
\usepackage[acronym]{glossaries}
\usepackage[hidelinks]{hyperref}
\usepackage{braket}
\usepackage{float}

\usepackage[dvipsnames]{xcolor}

\begin{document}

\title{Geometry-Driven Suppression of Fermionic Pairing on a Spherical Surface}

\author{L. Frigato}
\affiliation{Dipartimento di Fisica e Astronomia ``Galileo Galilei", Università di Padova, Via Marzolo 8, 35131 Padova, Italy}
\affiliation{Istituto Nazionale di Fisica Nucleare (INFN), Sezione di Padova, via Marzolo 8, 35131 Padova, Italy}

\author{A. Tononi}
\affiliation{Department de Física, Universitat Politècnica de Catalunya, Campus Nord B4-B5, E-08034, Barcelona, Spain}

\author{L. Salasnich}
\affiliation{Dipartimento di Fisica e Astronomia ``Galileo Galilei", Università di Padova, Via Marzolo 8, 35131 Padova, Italy}
\affiliation{Istituto Nazionale di Fisica Nucleare (INFN), Sezione di Padova, via Marzolo 8, 35131 Padova, Italy}
\affiliation{Istituto Nazionale di Ottica (INO) del Consiglio Nazionale delle Ricerche (CNR), via Nello Carrara 1, 50019 Sesto Fiorentino, Italy}

\date{\today}

\begin{abstract}
We investigate the ground-state properties of a two-component attractive Fermi gas confined to a spherical surface. In the strongly attractive regime, the system forms a Bose-Einstein condensate (BEC) of tightly bound dimers, yielding a gap and chemical potential that approach flat-space results with leading-order geometric corrections. 
As the attraction decreases, the system crosses over into a Bardeen-Cooper-Schrieffer (BCS) regime, the sphere radius $R$ constrains the Cooper pair size and shell effects emerge. For closed shells, corresponding to magic particle numbers set by the single-particle spectrum, pairing is suppressed above the scattering length $a^*=4R$, whose value is universal in the spherical geometry in contrast with box-trapped fermionic gases.
Our results highlight a fundamental interplay among geometry, quantum statistics, and interactions, motivating future experiments with shell-trapped ultracold fermions and inspiring geometry-controlled condensed matter devices.

\end{abstract}

\maketitle

The role of spatial geometry in physics has evolved through the centuries from the absolute passive container of Newtonian mechanics to the medium of gravitational interaction in Einstein’s general relativity \cite{jammer2013}. 
In contemporary research, control over the curvature of low-dimensional condensed-matter systems enables engineering their physical functionality and tuning their interactions \cite{CurvedRev2, CurvedRev3, CurvRev4, CurvRev5}.
Advances in the trapping of ultracold atoms now allow gases to be confined to thin, non-Euclidean geometries \cite{Revdubhelen}, raising fundamental questions about how spatial geometry \textit{per se} affects the properties of low-dimensional quantum systems.

Ultracold shell-shaped gases \cite{Lundblad_2023, Revdubhelen, Tononi2024} have become the main geometry for this kind of exploration, aiming to elucidate the role of spatial curvature, boundaries, and topology in quantum gases \cite{Tononi2023}.
Both experiments, which realize proposals \cite{Firstproposal, Lundblad2019, Wolf2022} for thin hollow shells with magnetic \cite{Firstrealization, Carollo2022, Perrin2022} or optical \cite{Jia2022, huang2025} potentials, and theoretical studies have focused mainly on the spherical geometry, which is the simplest curved surface, with no boundaries and nontrivial topology. These features modify both the equilibrium \cite{TononiBECsphere, bereta2019, rhyno2021, Biral_2024, Moller_2020} and dynamical \cite{CurvedRev1, StaticdynBEC, Superfluidvortex, Rhyno2026} properties of the system, giving rise to qualitative corrections in the few- to many-body regimes \cite{Zhang_2018, Tononiscattering, DimeroTononi, topologicalpelster, Tononi2025} that have no flat-space counterparts. 
While most of these studies focused on bosons, far fewer have explored fermionic gases confined to spherical surfaces
\cite{Fermisphere3, Fermisphere2, Fermisphere1, Frigato2026, Fermisphere4}, pointing toward shell-trapped Fermi gases as the next research frontier.

In this Letter, we show how the spherical geometry affects pairing in a two-component Fermi gas with attractive interactions. 
Using a functional field theory formalism, we first show that the pairing gap and chemical potential exhibit geometry-induced deviations from planar Fermi gases. In particular, we obtain leading geometric corrections to planar results in the deep BEC limit, where fermions form tightly bound dimers that perturbatively feel the curved geometry. 
When moving towards the BCS regime, the increasing spatial extent of Cooper pairs is bounded by the radius $R$, and the underlying geometry qualitatively affects fermionic pairing. 
Shell effects emerge in this regime, and pairing is strongly suppressed for atomic numbers corresponding to closed-shell configurations. 
This suppression occurs at the scattering length value $a^*=4R$, which, in the spherical geometry, is a universal constant essentially independent of the particle number. 
This result contrasts the highly irregular pairing suppression pattern displayed in flat periodic boxes.

Our results can be observed in optically trapped Bose-Fermi mixtures such as $^{6}\text{Li}$-$^{23}\text{Na}$ \cite{LiNaexp}, or in other heteronuclear mixtures enabling co-located trapping of distinct species \cite{Safronova2006}. 
Specifically, a thin shell of two-component $^{6}\text{Li}$ atoms surrounding a bosonic $^{23}\text{Na}$ core can be realized by magnetically tuning the interactions.
This realization extends the core-shell approach previously used to produce shell-shaped Bose-Einstein condensates in $^{87}\text{Rb}$-$^{23}\text{Na}$ Bose-Bose mixtures \cite{Jia2022, huang2025, DajunWang, DajunWang2, DajunWang3}. 
Such an experiment would provide a highly controllable platform to investigate pairing and vortices in a curved geometry, also offering a potential simulator for superfluid shell physics in neutron stars \cite{Revpairnucl}.

\textit{Theoretical model.---} 
We consider a two-component Fermi gas of mass $m$ confined to a spherical surface of radius $R$.
Fermions are described by the Grassmann fields $\psi_{\sigma}$, with $\sigma \in \{\uparrow, \downarrow\}$ labeling the different species. 
Same species fermions are noninteracting, while different species attract each other via a zero-range two-body potential characterized by the bare coupling constant $-g < 0$. 
These interactions lead to the formation of Cooper pairs, described by the bosonic pairing field $\Delta$.
The system is governed by the Euclidean Lagrangian density \cite{Altland_Simons_2010}
\begin{equation}
\mathscr{L}\! =\! \sum_{\sigma}\! \bar{\psi}_\sigma \Big[ \hbar \partial_\tau + \frac{\hat{L}^2}{2mR^2} - \mu \Big] \psi_\sigma - \Delta^* \psi_\downarrow\psi_\uparrow - \bar{\psi}_\uparrow \bar{\psi}_\downarrow \Delta + \frac{|\Delta|^2}{g},
\label{Lagrangian density}
\end{equation}
where the chemical potential $\mu$ fixes the particle number of both species, and $\tau=it$ is the imaginary time.
The spherical geometry is encoded in the kinetic energy operator, expressed via the angular momentum operator squared $\hat{L}^2$, in the boundary conditions of the fields $\Delta$ and $\psi_{\sigma}$, and in their dependence on the polar and azimuthal angles $\theta \in [0,\pi]$ and $\varphi\in [0,2\pi]$.

Let us derive the thermodynamic properties of the attractive Fermi gas on the spherical surface.
Thermodynamics is encoded in the grand canonical potential $\Omega = -\beta^{-1} \ln Z$, obtained via functional integration of the partition function $Z = \int \mathcal{D}[\Delta,\psi] \ e^{-S/\hbar}$ \cite{Altland_Simons_2010}, where the Euclidean action reads $S=\int_0^{\hbar \beta} d\tau \int_0^{2\pi} d\varphi \int_0^\pi d\theta  \sin\theta  { R^2} \, \mathscr{L}$, with $\beta = (k_B T)^{-1}$ and $k_B$ the Boltzmann constant. 
Assuming a uniform and static pairing field $\Delta \in \mathbb{R}$, the action is invariant under spatial rotations and the Grassmann fields can be conveniently expanded in the single-particle basis of spherical harmonics $Y_{l,m_l}(\theta, \varphi)$.
These states are labeled by orbital angular momentum $l = 0, 1, 2,\dots$ and by the magnetic quantum number $m_l = -l, -l+1, \dots, l$, and, in this representation, Cooper pairs form between time-reversed states $(l, m_l, \uparrow)$ and $(l, -m_l, \downarrow)$. 
Integrating out the fermionic fields and summing over the fermionic Matsubara frequencies (see End Matter), we obtain the grand canonical potential 
\begin{equation}
 \Omega = 4\pi R^2\frac{\Delta^2}{g}+ \sum_{l=0}^{+\infty} (2l+1) \left[(\xi_l-E_l)-\frac{2}{\beta}\ln(1+e^{-\beta E_l})\right],
 \label{Grand canonical potential}
\end{equation}
where $E_l=\sqrt{\xi_l^2+\Delta^2}$ is the quasiparticle excitation spectrum, $\xi_l=\epsilon_l-\mu$, and $\epsilon_l = \frac{\hbar^2}{2mR^2} l(l+1)$ is the single-particle energy spectrum.
The following results specialize to the zero-temperature case, $\beta \to \infty$.

The grand canonical potential \eqref{Grand canonical potential} is a function of the chemical potential $\mu$ and of the pairing gap $\Delta$.
These parameters are determined by self-consistently solving two coupled equations.
The first one is the number equation
\begin{equation}
   n = \frac{1}{4\pi R^2}\sum_{l=0}^{+\infty} (2l+1) \left( 1- \frac{\xi_l}{E_l} \right),
\label{Number equation}
\end{equation}
where we fix the two-dimensional number density $n = N/(4\pi R^2)$ as an input, and $N = - \partial_{\mu} \Omega$. 
The second one is the gap equation  \footnote{Here we look for the nontrivial solution $\Delta\ne0$. However, $\Delta=0$ always exists and becomes relevant at closed shells, as shown below.}
\begin{equation}
\begin{aligned}
    \frac{1}{g} = \frac{1}{4\pi R^2}\sum_{l=0}^{+\infty}\frac{2l+1}{2E_l},
\end{aligned}
\label{gap1}
\end{equation}
obtained by extremizing the grand potential as $\partial_{\Delta} \Omega = 0$.
While the number equation converges rapidly, the summand of the gap equation scales as $1/l$ at large angular momenta $l$, a logarithmic ultraviolet divergence typical of zero-range interactions \cite{Legget}. This divergence is easily regularized by expressing the bare coupling $g$ in terms of the two-body binding energy $\varepsilon_B < 0$ through the relation
\begin{equation}
\frac{1}{g}=\frac{1}{4\pi R^2}\sum_{l=0}^{+\infty} \frac{2l+1}{2\epsilon_l-\varepsilon_B},
\label{two body eq state}
\end{equation}
which can be directly obtained from the two-body Schrödinger equation for the ground-state dimer \cite{Zhang_2018, Tononiscattering}.
Equating Eqs.~\eqref{gap1} and \eqref{two body eq state} cancels the ultraviolet divergence, 
yielding the regularized gap equation
\begin{equation}
\begin{aligned}
    \sum_{l=0}^{+\infty}(2l+1)\left(\frac{1}{2E_l} - \frac{1}{2\epsilon_l - \varepsilon_B} \right) = 0.
\end{aligned}
\label{Gap equation}
\end{equation}
The number equation \eqref{Number equation} and the regularized gap equation \eqref{Gap equation} thus form a closed set that numerically determines the chemical potential $\mu$ and the gap $\Delta$ as a function of the two-body binding energy $\varepsilon_B$.

Let us finally relate $\varepsilon_B$ to the experimentally controllable two-dimensional $s$-wave scattering length $a$, and to the sphere radius $R$.
Unlike in flat space, $\varepsilon_B$ depends on both variables simultaneously, requiring the two-body binding problem to be solved directly on the curved surface \cite{DimeroTononi, Tononi2025}. 
Their exact relation is given by 
\begin{equation}
\ln\left( \frac{R}{a}\right) = \frac{\mathcal{D}(1/2 + s) +
\mathcal{D}(1/2 - s)}{2} + \ln \left(\frac{e^{\gamma_E}}{2}\right) 
\label{twobodysolution}
\end{equation}
where $s = ( 1/4+m R^2 \varepsilon_B / \hbar^2 )^{1/2}$, $\mathcal{D}$ is the digamma function, and $\gamma_E=0.577 \dots$ is the Euler constant \cite{Tononi2025, DimeroTononi}. 
The binding energy admits the strong-coupling expansion $\varepsilon_B \simeq \bar{\varepsilon}_B - c_B/R^2$, valid when $a/R \ll 1$.
The first term recovers the standard flat-space result $\bar{\varepsilon}_B = -4\hbar^2 e^{-2\gamma_E}/(ma^2)$ for the two-dimensional binding energy \cite{landauQM}, while the second is a geometric correction $c_B = \hbar^2/(3m)$ that only vanishes as $R \to \infty$. 
Assuming $\varepsilon_B \sim \bar{\varepsilon}_B$ and replacing sums with integrals \cite{Fermisphere3} discards the distinctive geometric features of the sphere. 
Here we retain both the discrete spectrum, which already revealed shell effects in a weakly repulsive gas \cite{Frigato2026}, and the exact two-body solution \eqref{twobodysolution}. 
Neglecting them would not capture the interplay between pairing and geometry, especially occurring in the BCS regime $\ln(a/R) \gtrsim 0$.

\begin{figure}
\centering
\includegraphics[width=0.9\linewidth]{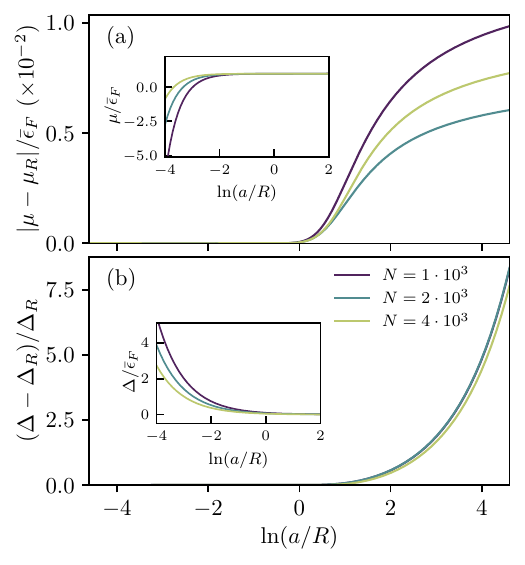}
\caption{Differences between the exact and low-curvature asymptotic values of: (a) the chemical potential, and (b) the gap, displayed versus the crossover parameter $\ln(a/R)$. 
The exact values of $\mu$ and $\Delta$ are obtained by solving Eqs.~\eqref{Number equation} and \eqref{Gap equation}, the asymptotic values of $ {\mu}_R$ and $ {\Delta}_R=\bar{\Delta}$ are given by Eq.~\eqref{EMexpansion}.
The observed convergence signifies that curvature provides quantitative corrections when fermions pair into tightly bound dimers of size $\sim a \ll R$ and, conversely, that beyond $\ln(a/R) \approx 0$ the system geometry strongly alters the flat-space results.
The insets display the values of the exact $\mu$ and $\Delta$ versus $\ln(a/R)$, and we choose here non-magic numbers of atoms $N$.
}
\label{fig1}
\end{figure}

\textit{Solution of the number and gap equations.---} 
The Fermi gas thermodynamics on the spherical surface is determined by the numerical solution of the coupled number and gap equations \eqref{Number equation} and \eqref{Gap equation}.
Inputs of the equations are the number density $n$ and the exact binding energy on the sphere $\varepsilon_B(a/R)$, provided by the solution of Eq.~\eqref{twobodysolution} for any given $s$-wave scattering length ratio $a/R$.
We obtain the numerical solution via a self-consistent procedure in which equations are alternately solved through a root-finding algorithm until convergence is reached. 

In the insets of Fig.~\ref{fig1}, we plot the values of $\mu/\bar{\epsilon}_F$ and $\Delta/\bar{\epsilon}_F$ versus $\ln(a/R)$ for partially filled shells \cite{Frigato2026}, where we use the flat-space Fermi energy $\bar{\epsilon}_F = \hbar^2 \pi n / m$ as energy scale.
Along the crossover from BEC ($a/R \ll 1$) to BCS ($a/R \gg 1$), the chemical potential crosses over from large negative values to positive ones and the gap smoothly vanishes, as in planar Fermi gases.

To gain insight into geometric effects, it is first instructive to obtain analytical approximations of Eqs. \eqref{Number equation} and \eqref{Gap equation} in the deep BEC regime $a/R \ll 1$. 
In this strongly attractive regime, the dimer has size $\sim a$, much smaller than the sphere radius \cite{SolitonsTononi}.
These tightly bound fermionic pairs are therefore confined to a slightly curved two-dimensional space, and planar results should be asymptotically reproduced.
To substantiate this intuition, we calculate the curvature corrections to the planar results by applying the Euler-Maclaurin summation formula to the discrete Eqs. \eqref{Number equation} and \eqref{Gap equation}. 
The equations expand to the planar-case continuous integral expressions plus finite-size corrections, and solving the expanded system consistently to order $\mathcal{O}(R^{-2})$ (see End Matter) yields
\begin{equation}
{ \mu_R} = \bar{\mu} + \frac{c_\mu}{R^2},
\quad {\Delta_R} = \bar{\Delta} + \frac{c_\Delta}{R^2}, 
\label{EMexpansion}
\end{equation}
where $\bar{\mu} = \bar{\epsilon}_{F} + \bar{\varepsilon}_B/{2}$ and $\bar{\Delta}= \sqrt{-2 \bar{\epsilon}_{F} \bar{\varepsilon}_B }$ recover the exact planar solutions. 
The geometric coefficients are $c_\mu=-\hbar^2/(6m)$ and $c_\Delta=0$, so that $\Delta_R=\bar{\Delta}$ up to $\mathcal{O}(R^{-4})$. Interestingly, $c_\mu$ is a constant independent of the interaction strength, meaning that the leading curvature correction is purely geometric.

We demonstrate the convergence of the full numerical solution to the asymptotic expressions \eqref{EMexpansion} in Fig.~\ref{fig1}.
For this demonstration, we take $N$ values corresponding to partially-filled non-interacting shells \cite{Frigato2026} and plot the difference between exact and asymptotic values: for the chemical potential, which changes sign across the crossover, we rescale by $\bar{\epsilon}_F$, whereas the gap is rescaled by $\Delta_R$.
Fig.~\ref{fig1} shows that the asymptotic expressions \eqref{EMexpansion} provide good approximations of the exact solution up to $\ln(a/R) \sim 0$, namely, as long as the dimer size is smaller than the radius $R$. 
Beyond this value, the solution must account for the correct two-body energy [which scales as $\varepsilon_B \propto -1/\ln (a/R)$ \cite{SolitonsTononi} and not as $\bar{\varepsilon}_B \propto - a^{-2}$ for $\ln(a/R) \gg 0$ \cite{Fermisphere3}], and retaining the discrete spectrum is necessary to observe the interplay between pairing and interactions in the BCS regime $\ln(a/R) \gtrsim 0$.

\begin{figure*}[hbtp]
\centering
\includegraphics[width=1.0\linewidth]{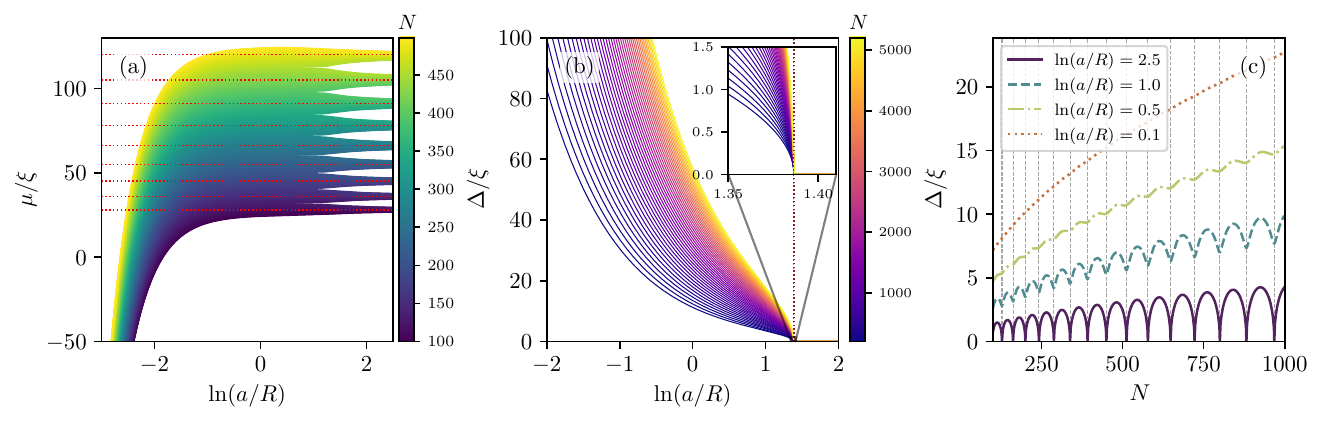}
\caption{Geometry-dependent shell effects in the BCS-BEC crossover on the surface of a sphere.
(a) The chemical potential $\mu/\xi$ versus $\ln(a/R)$ (excluding $N_{\rm magic}$ values), with $\xi = \hbar^2/(mR^2)$, shows the separation into bands and the convergence towards the closed-shell values when increasing $\ln(a/R)$ towards the weakly attractive BCS regime.
(b) The pairing gap $\Delta/\xi$ versus the interaction strength $\ln(a/R)$ for closed shells ($N = N_{\rm magic}$) displays a suppression below the interaction strength marked by the red dotted line.
The inset shows a zoom of the collapse region, which displays no dependence on $N$.
(c) Pairing oscillations of $\Delta$ are localized near $N_{\rm magic}$ values {(dashed vertical lines)}, and become more evident as $\ln(a/R)$ is increased. 
While strong attractions wash out the effects of discrete spectrum, weak interactions preserve distinct shell signatures.
}
\label{fig2}
\end{figure*}

\textit{Pairing suppression at the magic numbers.---} 
As interactions are decreased towards the weakly attractive BCS regime $a/R \gg 1$, shell effects characteristic of the spherical geometry become more visible.
This fact is demonstrated in Fig.~\ref{fig2}(a), where we plot $\mu/\xi$ as a function of $\ln(a/R)$ for different $N$ values, with $\xi = \hbar^2/(m R^2)$. As $\ln(a/R)$ increases and the gas moves towards the weakly attractive BCS regime, the chemical potential separates into distinct bands that converge to the noninteracting values $ \mu = \epsilon_F = \hbar^2 l_F(l_F+1)/(2mR^2)$ (red dotted lines), where we introduce the Fermi energy $\epsilon_F$ and the highest occupied shell $l_F = \lfloor \sqrt{(N-1)/2} \rfloor$, with $\lfloor ... \rfloor$ denoting the integer part. 

Such shell effects are of particular interest for magic numbers of atoms $N = N_{\rm magic}=2(l_F+1)^2$, corresponding to completely filled shells in the noninteracting system. 
In this case, Eqs.~\eqref{Number equation} and \eqref{Gap equation} predict that the system does not undergo a smooth crossover. 
This result is illustrated in Fig.~\ref{fig2}(b), where we plot the gap $\Delta/\xi$ versus $\ln(a/R)$ for different $N_{\rm magic}$ values. 
We find that the gap quickly decreases towards zero at the finite value $\ln(a^{*}/R) \simeq 1.39$, independent of $N_{\rm magic}$, signaling a strong suppression of fermionic pairing for weak attractive interactions. 
Complementary evidence of this suppression is provided by Fig.~\ref{fig2}(c), showing clear dips of $\Delta$ versus $N$ as $\ln(a/R)$ is increased and as $N \sim N_{\rm magic}$ \cite{Urban2003}.

The geometric suppression of pairing observed is explained by a simple physical argument, which also applies to finite Fermi systems with discrete spectra such as ultrasmall metallic grains \cite{Anderson1959, vonDelft2001} and harmonically trapped Fermi gases \cite{Heiselberg2002, Urban2003}.
For very weak attractions $\ln(a/R) \gg 1$, pairing occurs in a shell of width $\Delta$ about the Fermi surface, identified by the angular momentum $l_F$ and energy $\epsilon_F$ \cite{Frigato2026}. 
For $N \neq N_{\rm magic}$, the shell is partially filled, and Cooper pairs can form as highly correlated states sustained by scattering near $l_F$.
However, the shell is full for $N = N_{\rm magic}$, and since intra-shell scattering is forbidden by Pauli blocking, pairing is only possible by promoting fermions to the first available empty shell $l_F+1$. This transition requires overcoming a discrete kinetic energy cost equal to the shell gap $\delta_F=\epsilon_{l_F+1}-\epsilon_{l_F} = \sqrt{N/2}\, \hbar^2/(mR^2)$, which can only occur for pairing energies $\Delta \sim \delta_F$. 
Substituting in this expression the asymptotic gap $\bar{\Delta}= \sqrt{-2 \bar{\epsilon}_{F} \bar{\varepsilon}_B }$ yields the pairing suppression value $\ln(a^*/R) = \ln 2 - \gamma_E \sim 0.12$, which provides an order of magnitude estimate of the scattering length value above which pairing is unfavorable. 
Although qualitative, this estimate correctly predicts that the pairing suppression value is $N$-independent in the spherical geometry.

To quantitatively determine the onset of this pairing suppression, we analyze the stability of the normal state, characterized by $\Delta=0$. 
At a magic atom number $N=N_{\rm magic}$, the normal state leaves the chemical potential strictly unconstrained within the finite shell gap $(\epsilon_{l_F},\epsilon_{l_F+1})$. 
To uniquely fix the characteristic chemical potential $\mu^*$ at which the pairing suppression occurs, we demand that the particle number remains exactly $N_{\rm magic}$ as the gap $\Delta \to 0$. Expanding the number equation to leading order in $\Delta^2$ yields
\begin{equation}
\sum_{l\le l_F}\frac{2l+1}{(\epsilon_l - \mu^*)^2} = \sum_{l>l_F}\frac{2l+1}{(\epsilon_l-\mu^*)^2},
\label{eq:muc_main}
\end{equation}
which depends solely on the discrete shell structure via $l_F$ and allows determining $\mu^{*}$. 
The corresponding binding energy $\varepsilon_B^*$ at this point ($\Delta \to 0$, $\mu = \mu^*$) follows from the gap equation
\begin{equation}
\sum_{l=0}^{+\infty}\frac{2l+1}{2|\epsilon_l-\mu^*|} = \sum_{l=0}^{+\infty}\frac{2l+1}{2\epsilon_l - \varepsilon_B^*},
\label{eq:epsBc_main}
\end{equation}
which can be solved for $\varepsilon_B^*$.
Then, using Eq.~(\ref{twobodysolution}), we relate $\varepsilon_B^*$ to the value of $\ln(a^*/R)$ at which suppression occurs. 
Remarkably, this threshold can be analytically determined by converting the sums in \eqref{eq:epsBc_main} into combinations of digamma functions (see End Matter),
yielding
\begin{equation}
    \ln\left(\frac{a^*}{R}\right)=\ln 4+\frac{1}{32l_F^2}+\mathcal{O}\left( \frac{1}{l_F^3}\right),
\label{pairingthreshold}
\end{equation}
where corrections at $\mathcal{O}(l_F^{-2})$ are equivalent to $\mathcal{O}(N^{-1})$. 

\begin{figure}[t]
\centering
\includegraphics[width=1.0\linewidth]{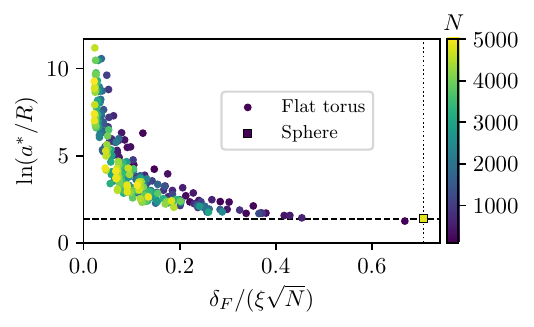}
\caption{Pairing suppression threshold $\ln(a^*/R)$ versus the rescaled shell gap $\delta_F/(\xi\sqrt{N})$ for closed-shell configurations on the sphere (square) and on a flat torus of equal area (circles), where points correspond to magic numbers $10^2<N_{\rm magic}<5\times 10^3$. 
The sphere point collapses at $\ln(a^*/R) = \ln 4$ (dashed line) and $\delta_F/(\xi\sqrt{N}) = 2^{-1/2}$ (dotted line), while flat torus points are widely distributed among $\ln(a^*/R)$ values that are inversely proportional to $\delta_F/(\xi\sqrt{N})$.
}
\label{fig3}
\end{figure}

The independence on the particle number of the leading order value $a^{*} = 4R$ is characteristic of the spherical geometry.
To prove it, we compare pairing in the sphere versus a flat torus of equal area (see End Matter). While both manifolds are compact and boundaryless, the sphere's rotational symmetry enforces a regular $2l+1$ shell degeneracy and a shell gap scaling $\delta_F \propto \sqrt{N}$, while the flat torus features irregular degeneracies and shell gaps with an erratic $N$ dependence (cf. Fig.~\ref{fig4}). 
Therefore, while the spherical geometric structure sets the pairing suppression at the universal threshold $a^*=4R$, pairing thresholds of the torus are highly irregular.
To verify this fundamental geometric distinction, we plot in Fig.~\ref{fig3} the suppression threshold $\ln(a^*/R)$ versus the rescaled shell gap $\delta_F/(\xi\sqrt{N})$ for magic numbers of the sphere (squares) and of the torus (circles). 
While the spherical points collapse to a single universal point, the flat torus pairing thresholds scale inversely with $\delta_F/(\xi\sqrt{N})$, since weaker attractions are sufficient to establish pairing for smaller shell gaps.

We finally remark that pairing suppression is obtained within a mean-field theory that models $\Delta$ as uniform and constant, pairing time-reversed fermions $(l, m_l, \uparrow)$ and $(l, -m_l, \downarrow)$ which belong to the same angular momentum shell $l$, and for which $\Delta=0$ is an exact solution in the very weakly attractive regime. 
Correlations of the pairing field $\Delta$ would favor pairing between different angular momentum shells, provided that angular momentum is conserved during the process, and are expected to restore a small residual pairing in the BCS regime, while still displaying a strong suppression around $a^* = 4R$.

\textit{Conclusions.---} In this Letter, we have investigated the zero-temperature BCS-BEC crossover of a two-component Fermi gas confined to the surface of a sphere. 
In the deep-BEC regime, where the pair size is much smaller than the sphere radius, the gap and chemical potential recover flat-space values asymptotically, with leading geometric corrections that we quantitatively estimate. 
When entering the BCS side of the crossover, the finite size of the sphere increasingly constrains the spatial extent of the Cooper pairs, producing sizable, shell-resolved deviations from the flat-space predictions. 
For closed-shell configurations, the discreteness of the spectrum turns the BCS-BEC crossover from smooth to step-like, since the pairing order parameter is suppressed below a given interaction strength, set by the kinetic energy cost of promoting a pair across the shell gap. 
This suppression threshold becomes universal, $a^* = 4R$, for sufficiently large $N$.
This phenomenon establishes geometry as a control parameter for fermionic superfluidity in compact curved spaces, and provides principles for designing novel condensed matter devices.

\begin{acknowledgments}
\textit{Acknowledgments.} We thank M. Urban for useful suggestions. L.~F.~ thanks A. Ponticelli, M. Lanaro, and C. Vianello for useful discussions and suggestions and acknowledges the project “Frontiere Quantistiche” within the 2023 funding program ‘Dipartimenti di Eccellenza’ of the Italian Ministry of Universities and Research. A.T.~acknowledges funding by the European Union under the Horizon Europe MSCA programme via the project 101146753 (QUANTIFLAC), support by the Spanish Ministerio de Ciencia, Innovación y Universidades (grant PID2023-147469NB-C21, financed by MICIU/AEI/10.13039/501100011033 and FEDER-EU), and discussions within the COST Action SCALES (CA24139), supported by COST (European Cooperation in Science and Technology). L.F. and L.S.~
are partially supported by
Iniziativa Specifica “Quantum" of Istituto Nazionale di Fisica Nucleare (INFN).  
\end{acknowledgments}

\vspace{-3mm}

\bibliographystyle{apsrev4-2}
\bibliography{bibliography.bib}

\vspace{-2mm}

\appendix

\clearpage

\section{End Matter}

\paragraph{Functional field integration.---}
We illustrate how to calculate the grand canonical potential $\Omega$ via functional integration.
The first step is diagonalizing the action $S$. 
This is achieved by expanding the Grassmann fields as
\begin{equation}
\psi_\sigma(\theta,\varphi,\tau) = \sum_{l=0}^\infty \sum_{m_l=-l}^l \sum_{n=-\infty}^{+\infty} { \frac{c_{l, m_l, \sigma}^{(n)}}{R}} e^{-i\omega_n \tau} Y_{l,m_l}(\theta,\varphi),
\end{equation}
where the spherical harmonics $Y_{lm}(\theta,\varphi)$ are the single-particle problem eigenfunctions, and Matsubara frequencies read $\omega_n=\frac{(2n+1)\pi}{\hbar\beta}$.
Using the symmetry property $Y_{l,m_l}^{*}(\theta, \varphi) = (-1)^{-m_l} Y_{l,-m_l}(\theta, \varphi)$, defining the Nambu spinor $\Psi_{l,m_l,n} = (c_{l, m_l, \uparrow}^{(n)}, \bar{c}_{l, -m_l, \downarrow}^{(-n)})^T$ of the expansion coefficients, and integrating out the spatial part of the action, we get the action in angular momentum space
\begin{equation}
S = \hbar \beta \sum_{l,m_l,n} \bar{\Psi}_{l,m_l,n} \hat{M}_{l,m_l,n} \Psi_{l,m_l,n} + \frac{4\pi { R^2}\hbar\beta\Delta^2}{g},    
\end{equation}
where the inverse Green's function matrix is given by
\begin{equation}
\hat{M}_{l,m_l,n} = \begin{pmatrix} -i\hbar\omega_n + \xi_l & -(-1)^{m_l}\Delta \\ -(-1)^{m_l}\Delta & -i\hbar\omega_n - \xi_l \end{pmatrix},
\end{equation} 
and the eigenvalues of $\hat{M}_{l,m_l,n}$ yield the BCS excitation spectrum $E_l=\sqrt{\xi_l^2+\Delta^2}$.
The partition function is then evaluated by performing the functional integration  
\begin{align}
     Z=e^{-4 \pi { R^2} \beta \frac{\Delta^2}{g}} \prod_{l=0}^{\infty} \prod_{m=-l}^{l} \prod_{n=-\infty}^{+\infty}\det\hat{M}_{l,m_l,n},
\end{align}
and the grand canonical potential is thus obtained as
\begin{equation}
    \Omega=4\pi{ R^2}\frac{\Delta^2}{g}-\frac{1}{\beta} \sum_{l=0}^{\infty} \sum_{m=-l}^{l} \sum_{n=-\infty}^{+\infty}\ln(\det\hat{M}_{l,m_l,n}).
\end{equation}
Known techniques \cite{Dupuis2023, Altland_Simons_2010} allow us to perform the Matsubara summation over $n$ and yield Eq.~\eqref{Grand canonical potential}.

\paragraph{Curvature correction in the BEC regime.---}
The leading-order expansion in $1/R$ of the spherical Eqs.~\eqref{Number equation} and \eqref{Gap equation} analytically reproduces the flat-space number and gap equations, yielding the planar solutions $\mu = \bar{\mu}$ and $\Delta = \bar{\Delta}$. Curvature corrections to these results are then obtained by using the Euler-MacLaurin formula,
\begin{equation}
\begin{aligned}
 \sum_{i=0}^{l_c} f(i) &= \int_{0}^{l_c} f(t)dt + \frac{1}{2}\left(f(l_c) + f(0)\right) +\\ &\sum_{s=1}^{k} \frac{B_{2s}}{(2s)!} \left(f^{(2s-1)}(l_c) - f^{(2s-1)}(0)\right) + R_k(f),
\end{aligned}
\label{Euler MacLaurin formula}
\end{equation}
where $B_{2s}$ are Bernoulli numbers (with $B_{2s+1} = 0$ for every $s \geq 1$), $k$ is a positive integer, $R_k$ is an error term, and $l_c \to +\infty$ at the end of the calculations. 
We use this formula to expand Eqs.~\eqref{Number equation} and \eqref{Gap equation} at $O(1/R^2)$, obtaining the curvature-corrected number and gap equations
\begin{align}
&n=\frac{m}{2\pi\hbar^2 }(\mu+\sqrt{\mu^2+\Delta^2}) \left( 1+ \frac{\hbar^2}{6m R^2}\frac{1}{\sqrt{\mu^2+\Delta^2}}\right),
\label{numberexpansion}
\\
&\ln \left({ \frac{ \varepsilon_B}{\mu-\sqrt{\mu^2+\Delta^2}}}\right)+ \frac{\hbar^2}{3 m R^2}\left[\frac{1}{2\sqrt{\mu^2+\Delta^2}}{+\frac{1}{\varepsilon_B} }\right]=0.
\label{gapexpansion}
\end{align}
By exponentiating Eq.~\eqref{gapexpansion}, extracting $\varepsilon_B$, and further expanding at $O(1/R^2)$, we obtain
\begin{equation}
\begin{aligned}
\varepsilon_B= {(\mu-\sqrt{\mu^2+\Delta^2})}\left[1- \frac{\hbar^2}{3mR^2} \left(\frac{1}{2\sqrt{\mu^2+\Delta^2}}{+\frac{1}{\varepsilon_B} }\right) \right].
\end{aligned}
\label{gapexpansion2}
\end{equation}
We then substitute the expansions $\mu \to { \mu_R}$ and $\Delta \to { \Delta_R}$ of Eq.~\eqref{EMexpansion} into Eqs.~\eqref{numberexpansion} and \eqref{gapexpansion2}, use $\varepsilon_B = \bar{\varepsilon}_B - c_B/R^2$, and solve for $c_\mu$ and $c_\Delta$ to obtain the values which appear in Eq.~\eqref{EMexpansion}.

\textit{Pairing suppression universality.---} 
Here we determine analytically the pairing suppression threshold $\ln(a^*/R)$ by solving analytically Eqs.~\eqref{eq:muc_main} and \eqref{eq:epsBc_main} in the limit of large $N$, equivalent to large $l_F$.
We first rewrite Eq.~\eqref{eq:epsBc_main} by including a large angular momentum cutoff $L \gg l_F$ in the sums, obtaining
\begin{equation}
\sum_{l=l_F+1}^{L} f_l(c) - \sum_{l=0}^{l_F} f_l(c) = \sum_{l=0}^{L} f_l(s),
\label{middle step}
\end{equation}
where we introduce the parameters $c = (2{\mu}^*/\xi+ 1/4)^{1/2}$ and $s$ defined after Eq.~\eqref{twobodysolution}, and $f_l(x) = \frac{2l+1}{(l+1/2)^2 - x^2} = (l+1/2+x)^{-1} + (l+1/2-x)^{-1}$. 
Using the property $\sum_{l=p}^{q}(l+y)^{-1} = \mathcal{D}(y+1+q) - \mathcal{D}(y+p)$, and expanding the digamma functions as $\mathcal{D}(L+y)\sim_{L \to \infty} \ln L$, we obtain 
\begin{equation}
\begin{aligned}
\sum_{j=\pm1} \Bigg[ \mathcal D\left(\frac{1}{2}+jc\right) -2 \mathcal D\left(\frac{3}{2}+jc+l_F\right)\\ + \mathcal D \left(\frac{1}{2}+js\right) \Bigg] = 0,
\label{Digammaexact}
\end{aligned}   
\end{equation}
where we recognize in the last term the combination of digamma functions that appears in Eq.~\eqref{twobodysolution}.
We substitute them into Eq.~\eqref{Digammaexact}, then assume that Eq.~\eqref{eq:muc_main} is solved by the value $\mu^* = (\epsilon_{l_F+1}+\epsilon_{l_F})/2$ in the middle of the shell gap at $l_F \gg 1$, corresponding to $c = l_F + 1 + \mathcal{O}(1/l_F)$.
This yields the pairing suppression threshold obtained in Eq.~\eqref{pairingthreshold}.

\paragraph{Pairing suppression in the flat torus geometry. ---}  
To elucidate the peculiarity of pairing in the spherical geometry, we comparatively solve the analogous number and gap Eqs.~\eqref{Number equation} and \eqref{Gap equation} in the flat torus geometry. 
We parametrize this flat boundaryless surface by coordinates $\mathbf{r}$ spanning the two-dimensional interval $[-L/2, L/2]^2$ with periodic boundary conditions, where $L=2\sqrt{\pi}R$. 
In this geometry, the number and gap equations are obtained by performing the following substitutions in Eqs.~\eqref{Number equation} and \eqref{Gap equation}: $E_l \to E_k=\sqrt{\xi_k^2+\Delta^2}$, $\xi_l \to \xi_k=\epsilon_k-\mu$ and $\epsilon_l \to \epsilon_k=\hbar^2 k^2/(2m)$, where momenta read $\mathbf{k}=2\pi\mathbf{n}/L$ and the sums are performed over $\mathbf{n}=(n_x,n_y) \in \mathbb{Z}^2$. 
The gap equation also contains the flat torus dimer energy $\varepsilon_B$, 
which we calculate by solving the relative Schrödinger equation $-\frac{\hbar^2}{m}\nabla^2
\psi(\mathbf{r})=\varepsilon_B\psi(\mathbf{r})$, with wave function periodicity $\psi(\mathbf{r}+\mathbf{n}L)=\psi(\mathbf{r})$ and Bethe-Peierls boundary conditions $\psi(r \to 0) \propto \ln(r/a)$.
The solution, obtained by specializing the framework of Ref.~\cite{Hammer} to the case of contact interactions, reads:
\begin{equation}
\psi(\mathbf{r})\propto\sum_{\mathbf{n}} K_0\left(\frac{\sqrt{-\varepsilon_Bm}}{\hbar} |\mathbf{r}+\mathbf{n}L|\right),
\label{Torus wf}
\end{equation}
and it essentially consists of the sum of the infinite-plane relative solutions, repeated over all periodic copies of the flat torus.
Imposing the Bethe-Peierls boundary conditions yields the implicit relation between the scattering length $a$ and the binding energy $\varepsilon_B$
\begin{equation}
\varepsilon_B=\bar\varepsilon_B \,
\exp \left[2\sum_{\mathbf{n}\neq\mathbf{0}}
K_0\!\left(\frac{\sqrt{-m\varepsilon_B}}{\hbar}\,L\,|\mathbf{n}|\right)\right],
\label{eq: torus binding}
\end{equation}
which provides finite-size corrections to the planar asymptotic result $\varepsilon_B=\bar\varepsilon_B$, reproduced for $L \to \infty$.

\begin{figure}[hbtp]
    \centering
    \includegraphics[width=0.9\linewidth]{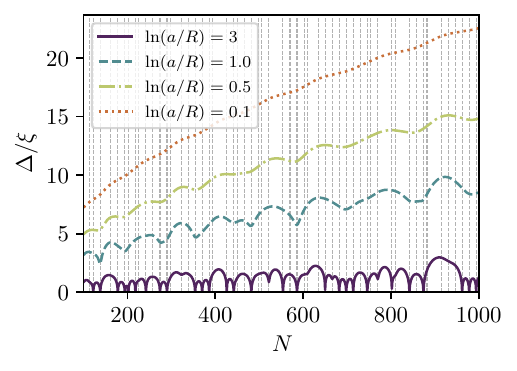}
    \caption{Pairing gap $\Delta/\xi$ versus the particle number for the flat torus geometry. 
    In contrast with the spherical geometry [cf. Fig. \ref{fig2}(c)], the flat torus magic numbers (dashed vertical lines) display no regular structure, and fermionic pairing suppression does not occur for all shells simultaneously at the same interaction value.}
    \label{fig4}
\end{figure}

Solving the flat torus number and gap equations using Eq.~\eqref{eq: torus binding} reveals a clear contrast with the spherical geometry. 
Indeed, as demonstrated by the comparison between Fig.~\ref{fig2}(c) and Fig.~\ref{fig4}, the flat torus exhibits an irregular shell structure governed by integer sums of squares ($n_x^2 + n_y^2 = n$), unlike the sphere's regular $2l+1$ level degeneracy.
Consequently, the torus shell gaps fluctuate erratically and lack the $\sqrt{N}$ scaling of spherical gaps. 
This implies that, at a fixed interaction strength, pairing may collapse at certain magic numbers while surviving at others, preventing the universal pairing suppression observed for the sphere.
This comparison demonstrates that the $N$-independent pairing collapse we observe is not a generic finite-size effect but is a direct signature of the sphere constant curvature and rotational symmetry.

\end{document}